\documentclass[twocolumn,showpacs,preprintnumbers,amsmath,amssymb]{revtex4}

\usepackage{graphicx}
\usepackage{dcolumn}
\usepackage{bm}

\begin{document}

\preprint{}
\input{epsf.tex}

\epsfverbosetrue

\title{Superradiant and Dark Exciton States in an Optical Lattice within a Cavity}
\author{Hashem Zoubi, and Helmut Ritsch}

\affiliation{Institut fur Theoretische Physik, Universitat Innsbruck, Technikerstrasse 25, A-6020 Innsbruck, Austria}

\date{05 May, 2009}

\begin{abstract}
We study ultracold atoms in a finite size one-dimensional optical lattice prepared in the Mott insulator phase and commonly coupled to a single cavity mode. Due to resonance dipole-dipole interactions among the atoms, electronic excitations delocalize and form {\it excitons}. These exciton modes are divided into two groups: antisymmetric modes which decouple from the cavity mode forming {\it dark states}, and symmetric modes significantly coupled to the cavity mode called {\it bright states}. In typical setups the lowest and most symmetric exciton is coupled to the cavity photons much stronger than the other bright states and dominates the optical properties response of the atoms ({\it superradiant state}). In the strong coupling regime this superradiant state is coherently mixed with the cavity photon to form a doublet of polariton states with the Rabi splitting.
\end{abstract}


\maketitle

The strong coupling regime of a single atom to a single cavity QED mode has been achieved experimentally in many different setups \cite{Haroche}, and even more extensively been studied theoretically \cite{Jaynes}. As recent breakthrough a large Bose-Einstein Condensate (BEC) within a cavity has been successfully used to demonstrate strong atom-field coupling \cite{Brennecke}. Experiments now even target to use BECs coupled to microwave atom chips \cite{Verdu,Imamoglu}, where strong coupling is expected even for magnetic collective coupling on dipole forbidden long lived hyperfine transitions. Theoretically, in a first approximation the coherent coupling of $N$ atoms of a BEC to a single cavity mode, with identical single atom-photon coupling $f$, is well described by the Tavis-Cummings model, which leads to a collective coupling of $\sqrt{N}f$ \cite{Tavis}. In this model the direct electrostatic interactions among the atoms are neglected. From the other side, a Bose gas of ultracold atoms loaded on an optical lattice realizes the quantum phase transition from the superfluid to the Mott insulator \cite{Greiner}, which is predicted by the Bose-Hubbard model \cite{Jaksch}. Moreover, controlling dipole-dipole shifts in an optical lattice clock was studied in \cite{Lukin}. These facts bring the fabrication of optical lattice ultracold atoms within a cavity very close. Such new set-up is of importance for light-matter interface in quantum information processing \cite{Simon}, and many-particle physics of quantum liquids \cite{Horak,Lewenstein}. This system exhibits bright and dark collective electronic excitations, which promise useful applications for quantum communication \cite{Pellizzari} and quantum memory \cite{Duan}.

Optical lattices are formed by prescribed counter propagating off resonant laser beams interacting with atomic motion only in a conservative way. For that atoms in such a lattice we consider here only the ground state and a single excited state of the atomic electronic excitations. We assume that their optical lattice potentials to have minima at the same positions and the atoms are localized in the corresponding first Bloch bands. The optical lattice is placed between spherical cavity mirrors, as seen in figure (1), so that only a single cavity mode close to resonance to the previous atomic transition. The superfluid to Mott insulator quantum phase transition within a cavity is studied by us in \cite{ZoubiA}, where we confirmed the existence of the Mott insulator, but for deeper optical lattices. Here we concentrate in the case of one-dimensional optical lattice in the Mott insulator phase with one atom per site, and in the case of low density of excited atoms. Different from the Tavis-Cummings model for a BEC within a cavity, we include the resonance dipole-dipole interaction among the atoms. This interaction gives rise to the formation of collective electronic excitations, with a superradiant state which is strongly coupled to cavity photons. Previously we investigated two-dimensional optical lattices in planar cavities, and due to resonance dipole-dipole interactions in exploiting the lattice symmetry, we got in-plane propagating excitation waves (excitons), and within a cavity in the strong coupling regime we obtained cavity polaritons \cite{ZoubiB}. In the present case for finite one-dimensional optical lattice, we have standing wave excitons coupled to the fixed cavity mode.

We present first the cavity photons. Only the lowest cavity mode is close to resonance to the atomic transition. Hence we consider only the Gaussian beam between the mirrors (see figure 1), which is represented by the Hamiltonian $H_c=E_c\ a^{\dagger}a$, where $a^{\dagger}$ and $a$ are the creation and annihilation operators of a cavity photon with energy $E_c=h\nu_c$, respectively. The electric field operator along the waist \cite{Haroche}, which is along the optical lattice, is
\begin{equation}
\hat{E}(r,z=0)=i\sqrt{\frac{E_c}{2\epsilon_0V}}\ e^{-r^2/w_0^2}\ \left\{\bar{e}\ a-\bar{e}^{\ast}\ a^{\dagger}\right\},
\end{equation}
where $\bar{e}$ is the photon polarization unit vector, $w_0$ is the beam waist, and $V$ is the mode volume, which is given by $V=\pi w_0^2L/4$, with $L$ the distance between the cavity mirrors, as seen in figure (1). Here we neglect the finite line width of both the cavity and the excited atomic states, to be included later in the linear optical spectra.

\begin{figure}[h!]
\centerline{\epsfxsize=4.0cm \epsfbox{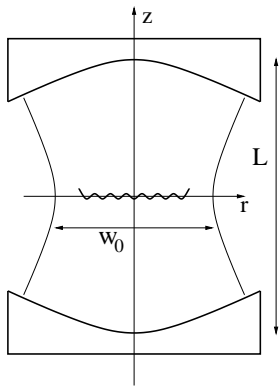}}
\caption{A cavity of spherical mirrors separated by distance $L$. We consider only the lowest Gaussian beam of waist $\omega_0$. The optical lattice forms of external lasers is located parallel to the waist.}
\end{figure}

Next we present electronic excitations in the optical lattice atoms. An electronic excitation at an atom site can transfer among the lattice atoms due to dipole-dipole resonance interactions. The excitation Hamiltonian is
\begin{equation}
H_{ex} = \sum_nE_a\ B_n^{\dagger}B_n + \sum_{\langle n,m\rangle}J_{\theta}\ B_n^{\dagger}B_m,
\end{equation}
where $B_n^{\dagger}$ and $B_n$ are the creation and annihilation operators of an electronic excitation at site $n$ with transition energy $E_a=h\nu_a$, respectively. At low number of electronic excitations we treat only a single excitation at a time. Namely the operators $B_n$ can be assumed to behave as bosons. We take into account only the interaction between nearest neighbor sites. The energy transfer coupling parameter is $J_{\theta}=\frac{\mu^2}{4\pi\epsilon_0a^3}\left(1-3\cos^2\theta\right)$, where $\mu$ is the atomic transition dipole, which makes an angle $\theta$ with the lattice direction. For isotropic atoms $\theta$ is fixed by the cavity photon polarization direction.

We consider a finite lattice with $N$ sites. The sites are labeled by $n=1,\cdots,N$. In order to use fixed boundary condition, we add two additional empty sites, $n=0$ and $n=N+1$, as in figure (2). Therefore the above excitation Hamiltonian can be diagonalize in using the transformation $B_n=\sqrt{\frac{2}{N+1}}\sum_k\sin\left(\frac{\pi n}{N+1}k\right)\ B_k$, where the collective excitation modes are labeled by $k=1,\cdots,N$, which are considered as standing excitons, and are presented in figure (2). The diagonal Hamiltonian reads $H_{ex}=\sum_kE_k\ B_k^{\dagger}B_k$, with the energy dispersion $E_k=E_a+2J_{\theta}\ \cos\left(\frac{\pi k}{N+1}\right)$. In place of discrete atom levels we get an energy band of band width $4J_{\theta}$.

\begin{figure}[h!]
\centerline{\epsfxsize=5.0cm \epsfbox{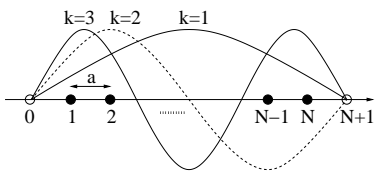}}
\caption{One-dimensional optical lattice of one atom per site Mott insulator, with lattice constant $a$. In using fixed boundary condition we get standing excitons. The first three exciton modes are plotted.}
\end{figure}

For the coupling between the electronic excitations and the cavity photons we use the dipole interaction $H_I=-\hat{\mu}\cdot\hat{E}$. The dipole operator is defined by $\hat{\mu}=\vec{\mu}\sum_n\left(B_n^{\dagger}+B_n\right)$, and we apply the rotating wave approximation. As the photon polarization at the waist is in the optical lattice direction, we choose here $\theta=0$ and then $\mu=\left(\bar{e}\cdot\vec{\mu}\right)$. The maximum of the Gaussian is taken in the middle of the optical lattice. Then the atom positions $r_n=na-\bar{L}/2$, with the lattice length $\bar{L}=(N+1)a$. For $w_0$ enough larger than $\bar{L}$, we use $e^{-r_n^2/w_o^2}\simeq 1$. We need also the summation over $n$, which is $\sum_n\sin\left(\frac{\pi n}{N+1}k\right)=\cot\left(\frac{\pi k}{2(N+1)}\right)$ for odd $k$-s, ($k=1,3,\cdots$), and zeros for even $k$-s, ($k=2,4,\cdots$). Finally we get $H_I=\sum_{k(odd)}\left(f_k\ B_k^{\dagger}a+f_k^{\ast}\ a^{\dagger}B_k\right)$, where $f_k=-i\sqrt{\frac{h\nu_c\mu^2}{\epsilon_0V(N+1)}}\ \cot\left(\frac{\pi k}{2(N+1)}\right)$.

The excitonic states are divided into two parts. The antisymmetric states of even $k$-s, ($k=2,4,\cdots$), with odd number of nodes. These states are decouple to the cavity photons, and are considered as dark states. The symmetric states of odd $k$-s, ($k=1,3,\cdots$), with even number of nodes. These states are coupled to cavity photons, and are considered as bright states. Here we show that the first mode $(k=1)$, which has no nodes, is a superradiant state, and the rest states ($k=3,5,\cdots$), are weakly coupled to the light. The things are clear due to the function $\cot\left(\frac{\pi k}{2(N+1)}\right)$ which decays very fast for small $k$-s, where $|f_1|/|f_{k\neq 1}|=k$. The oscillator strength is proportional to $|f_k|^2$, hence the oscillator strength of the first mode $(k=1)$ is stronger by $k^2$ from the $(k\neq 1)$ state, for small $k$-s. For example the oscillator strength of the first state is stronger $9$ times from the third state. As the oscillator strengths are summed to one, we conclude that, for $(N\gg 1)$, the first state $(k=1)$ includes $0.81$ of the sum, and the second bright state $(k=3)$ includes $0.09$.

To present the above results we consider an optical lattice within a cavity for the following numbers. The lattice constant is $a=10^{-7}\ [M]$, the beam waist is $w_0=3\times 10^{-4}\ [M]$, the distance between the mirrors is $L=1.5\times 10^{-3}\ [M]$, and the cavity mode volume is $V=10^{-10}\ [M^3]$. The atomic transition dipole is $\mu=5\times 10^{-29}\ [CM]$, the angle between the dipole and the optical lattice direction at the waist is taken to be $\theta=0$, and which results in the energy transfer parameter of $J_0/h=-6.8\times 10^7\ [Hz]$. The atomic transition frequency is $\nu_a=4\times 10^{14}\ [Hz]$. For the case of $N=10^3$ sites, in figure (3a) we plot the shifted exciton dispersion $E_k-E_a$ as a function of $k$. In figure (3b) we plot the square of the exciton-photon coupling $|f_k|^2$ as a function of $k$. Here the cavity photon frequency is taken to be at resonance with the first exciton mode, that is $E_c=E_1$. It is clear how the coupling decays very fast for large $k$. For the coupling of the first exciton we have $|f_1|/h=2.55\times 10^{7}\ [Hz]$, and for the third exciton we have $|f_3|/h=8.5\times 10^{6}\ [Hz]$.

\begin{figure}[h!]
\centerline{\epsfxsize=4.3cm \epsfbox{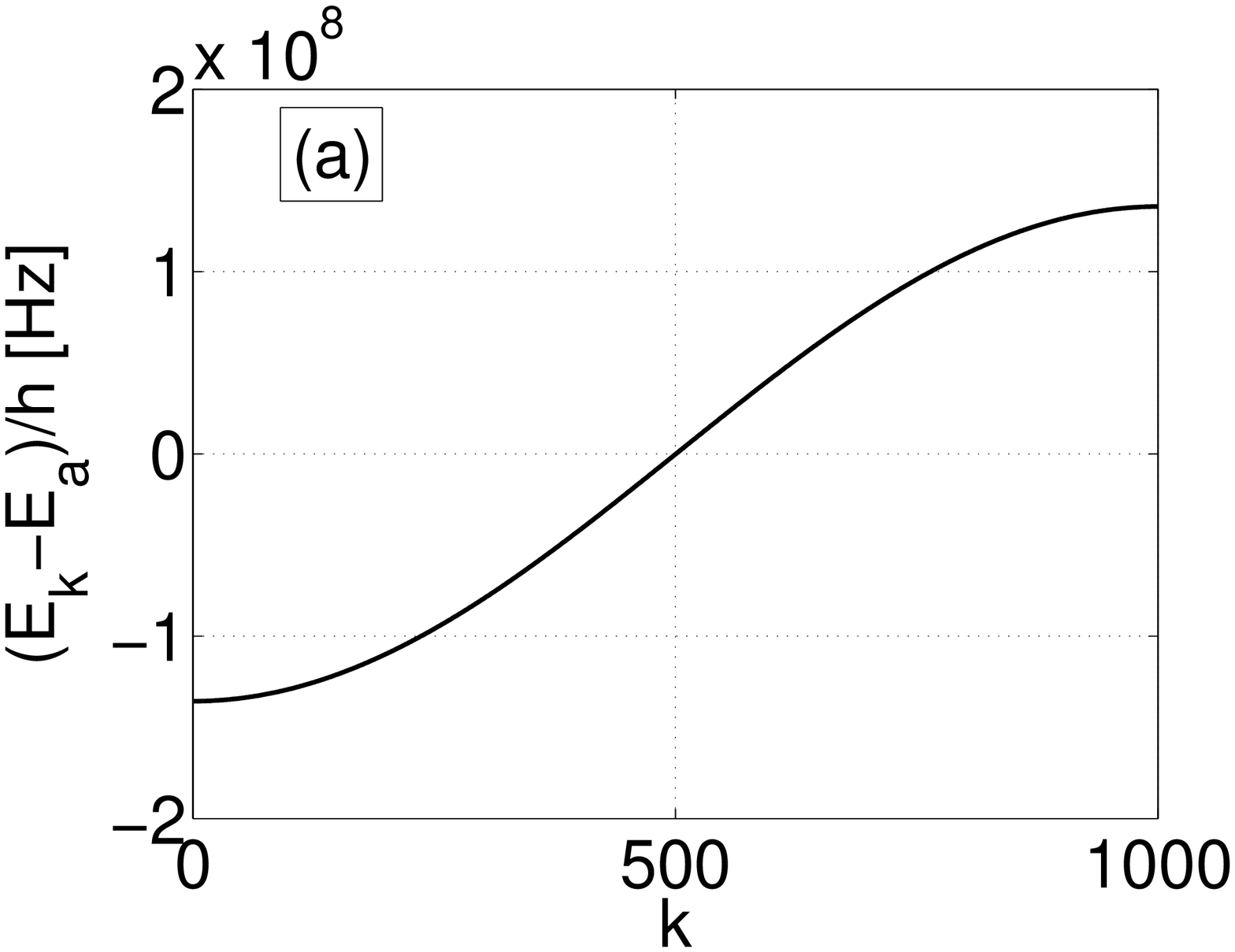}\epsfxsize=4.3cm \epsfbox{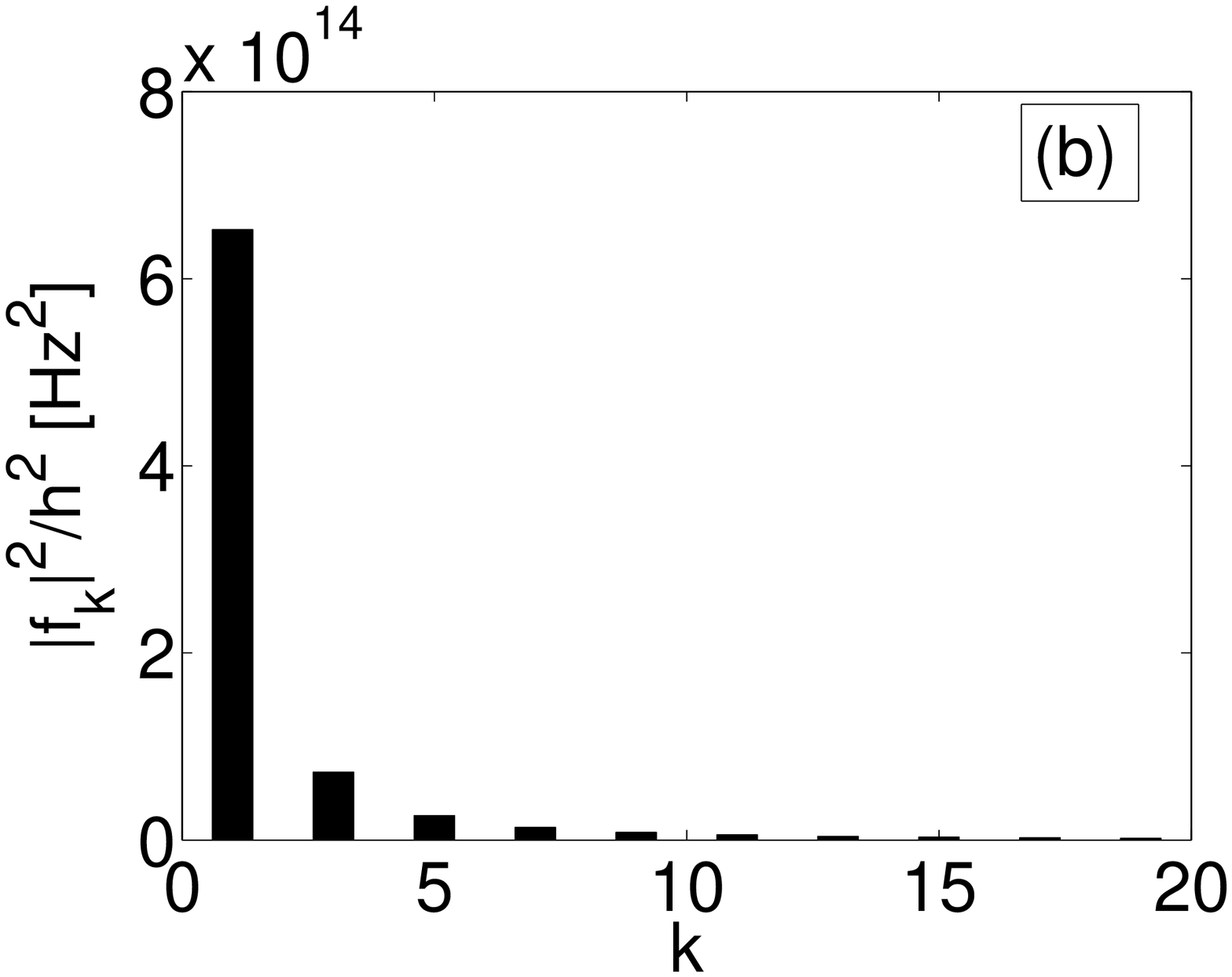}}
\caption{(a) The shifted exciton dispersion $(E_k-E_a)/h$ vs. $k$, for $N=10^3$ and $\theta=0$. The dispersion is a $cos$ function with band width $4J_0$. (b) The squared exciton-photon coupling $|f_k|^2/h$ vs. $k$, for $N=10^3$. It is seen that the first mode has the dominant coupling, which is nine times larger than the third mode. The even modes are dark.}
\end{figure}

In the light of the above discussion, we can neglect the coupling of the cavity photons to the exciton states except the coupling to the superradiant state. Therefore, the coupled photon-exciton Hamiltonian is
\begin{equation} \label{Hexciton}
H=E_{ex}\ B^{\dagger}B+E_c\ a^{\dagger}a+f\ B^{\dagger}a+f^{\ast}\ a^{\dagger}B,
\end{equation}
where we dropped the exciton index. The coupling parameter is
\begin{equation} \label{fexciton}
f=-i\sqrt{\frac{E_c\mu^2}{\epsilon_0V(N+1)}}\ \cot\left(\frac{\pi}{2(N+1)}\right),
\end{equation}
and the exciton energy is $E_{ex}=E_a+2J_{\theta}\ \cos\left(\frac{\pi}{N+1}\right)$. In the strong coupling regime, where the exciton-photon coupling is larger than their line widths, the Hamiltonian can be easily diagonalized in using the transformation $A_{\pm}=X^{\pm}\ B+Y^{\pm}\ a$, to get $H=\sum_r E_p^r\ A_r^{\dagger}A_r$, with the two polariton eigenenergies $E_p^{\pm}=(E_c+E_{ex})/2\pm\Delta$, where $\Delta=\sqrt{\delta^2+|f|^2}$, with the detuning $\delta=(E_c-E_{ex})/2$. The eigenstates are coherent superpositions of the exciton and the cavity photon. The exciton amplitudes are $X^{\pm}=\pm\sqrt{(\Delta\mp\delta)/2\Delta}$, and the cavity photon amplitudes are $Y^{\pm}=f/\sqrt{2\Delta(\Delta\mp\delta)}$. In figure (4a) we plot the shifted polariton dispersion $E_p^{\pm}-E_{ex}$ as a function of the detuning $\delta$, in using the previous numbers. In figures (4b) we plot the excitonic and photonic weights for the lower and upper branches.

\begin{figure}[h!]
\centerline{\epsfxsize=4.3cm \epsfbox{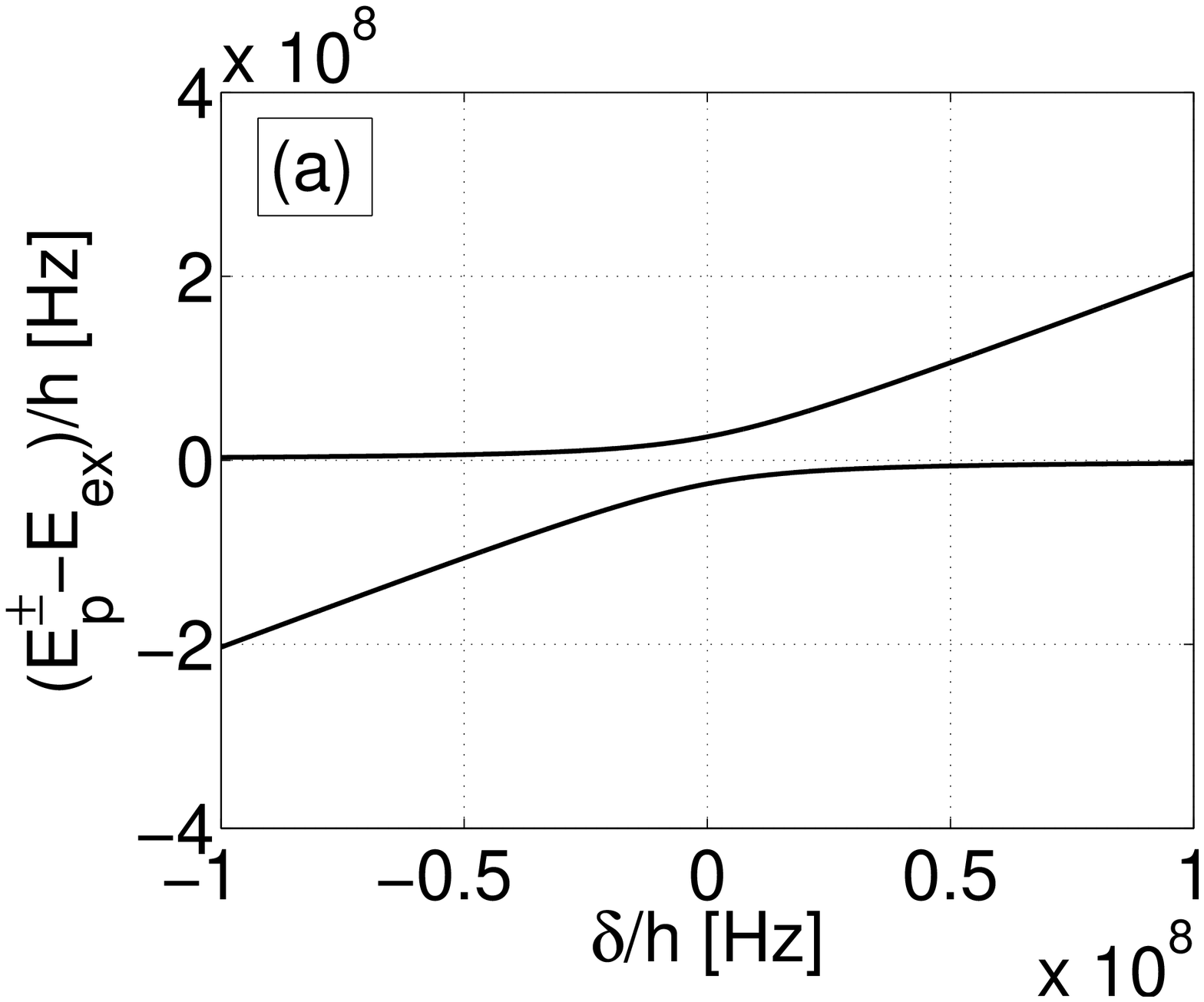}\epsfxsize=4.3cm \epsfbox{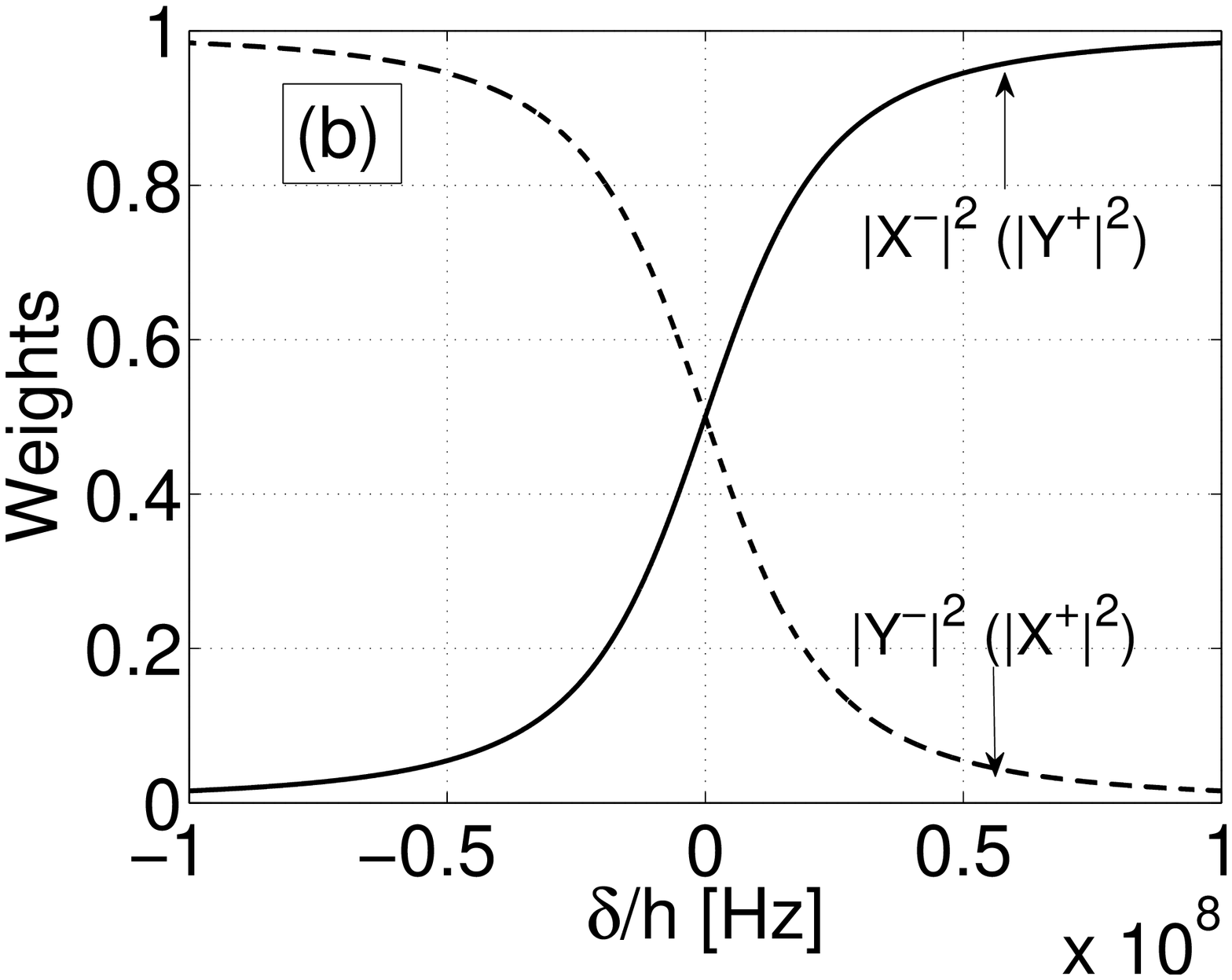}}
\caption{(a) The shifted polariton dispersions $(E_p^{\pm}-E_{ex})/h$ vs. the detuning $\delta/h$. The vacuum Rabi splitting is obtained at $\delta=0$. (b) The excitonic and photonic weights, $|X^{\pm}|^2,\ |Y^{\pm}|^2$, vs. the detuning $\delta/h$, for the lower and upper branches. At $\delta=0$ the branches are half exciton-half photon. For large positive detuning the lower branch becomes exciton and the upper becomes photon, and vice versa for negative detuning.}
\end{figure}

To observe the system we need to couple the cavity mode to the external radiation field. We do this by considering non-perfect cavity mirrors. We introduce cavity mode damping rates for the upper and lower mirrors by $\gamma$. As in experiments the cavity is open, we also include the cavity photon damping rate $\Gamma_c$ directly into the free space. The excited atoms have damping rate of $\Gamma_a$, which is included here phenomenologically. The external field serves us with two things: damping of the cavity mode, and input-output fields \cite{ZoubiB}.

We interest in the linear optical spectra. For an incident given field with a fixed polarization from the upper side of the cavity and normal to the mirror, in figure (5a) we plot the transmission spectra, and in figure (5b) the reflection spectra. We used the damping frequencies $\Gamma_a=\Gamma_c=\gamma=10^{7}\ [Hz]$. The two peaks and the two dips correspond to the two polariton states. We assumed here zero detuning, that is $\delta=0$, and $\theta=0$.

\begin{figure}[h!]
\centerline{\epsfxsize=4.3cm \epsfbox{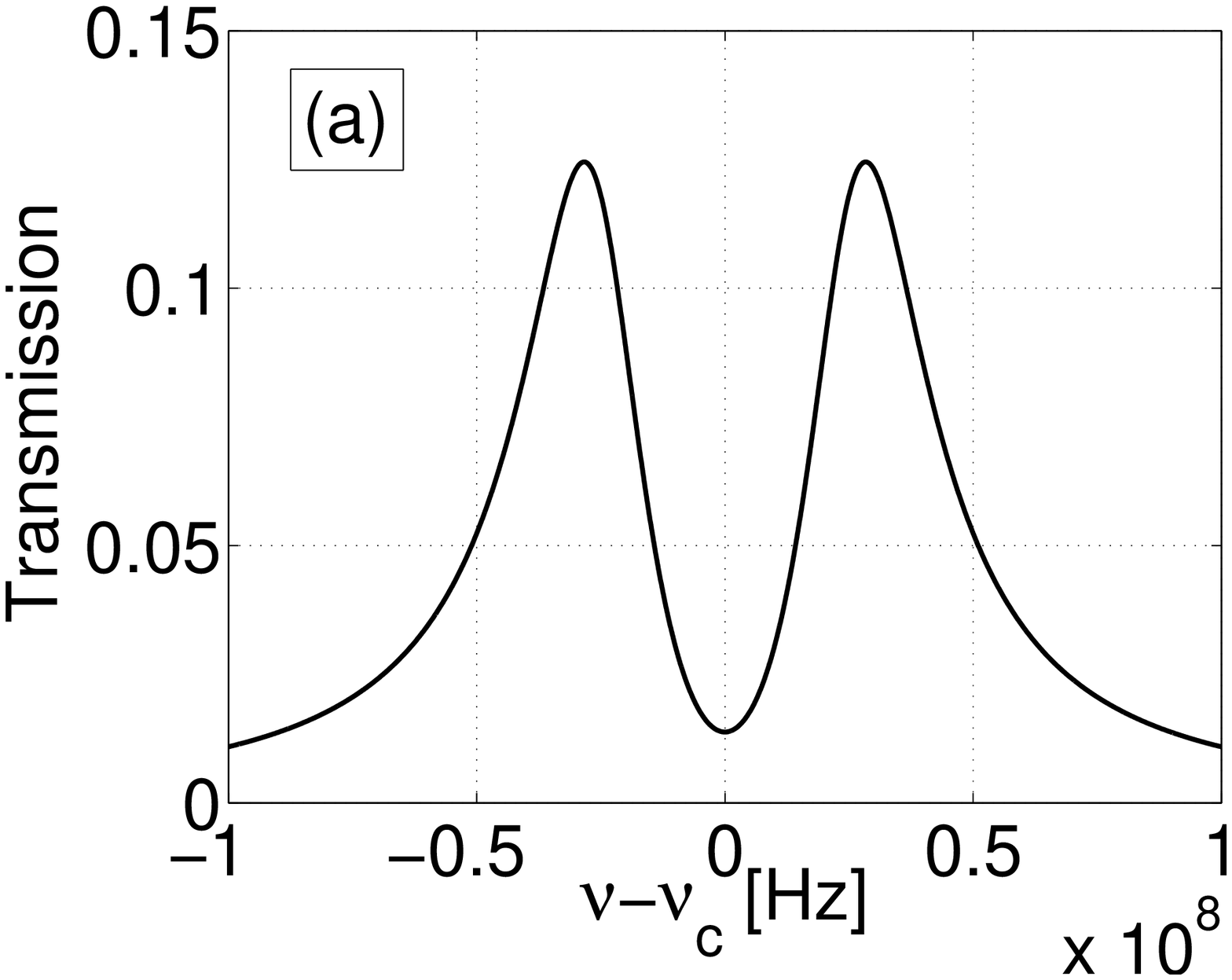}\epsfxsize=4.3cm \epsfbox{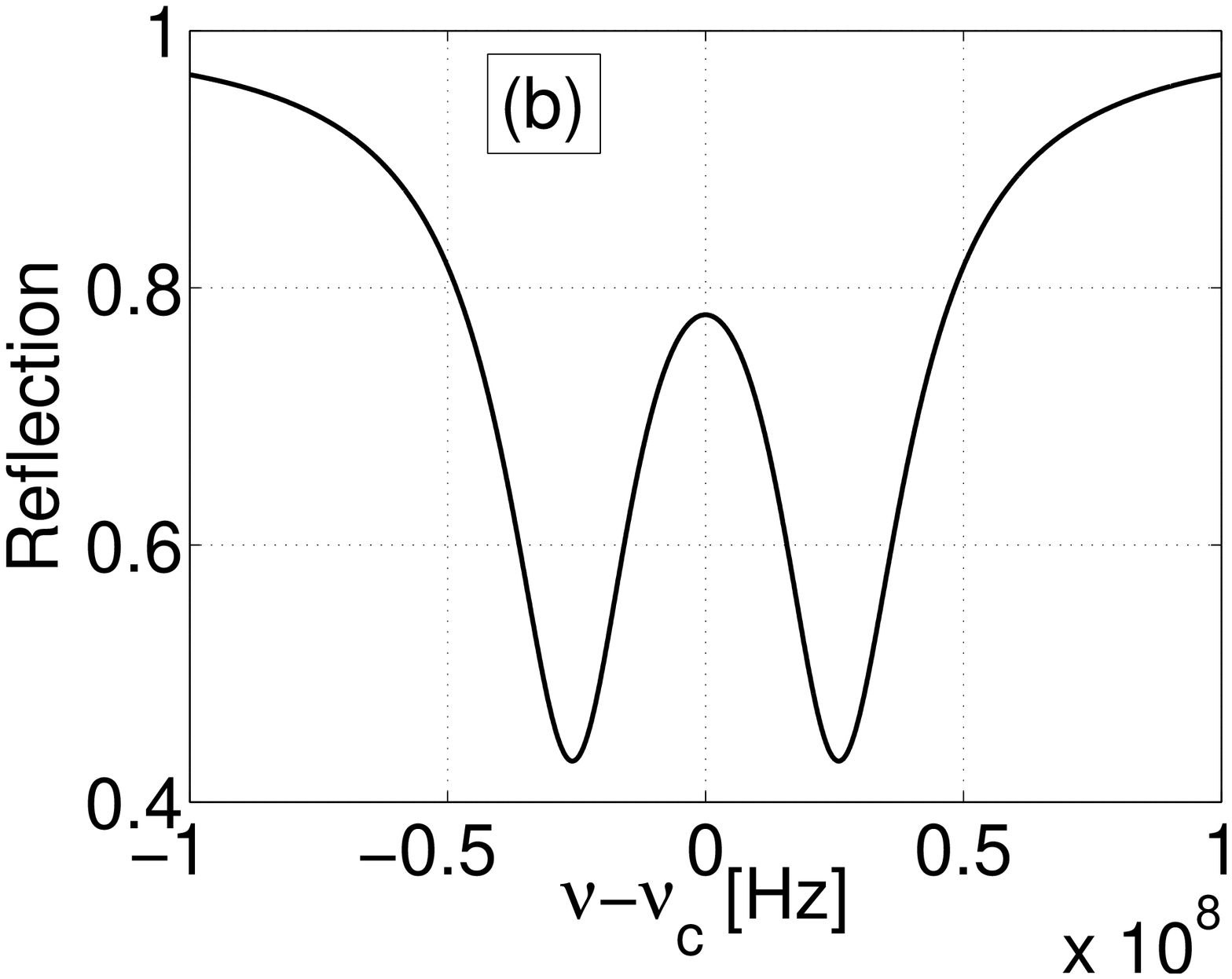}}
\caption{(a) The transmission spectra. (b) The reflection spectra. The two peaks and two dips correspond to the polariton doublet, at $\delta=0$. The spectrum line widths result of the cavity photon and excited atom line widths.}
\end{figure}

For comparison needs, we consider the same previous system, but now in neglecting the dipole-dipole interactions among the atoms. As we neglect also the spatial change of the cavity mode along the optical lattice, this makes the system similar to trapped ultraclod boson atoms within the previous cavity, in the middle between the mirrors, where the atoms are taken to be independent by neglecting the dipole-dipole interactions \cite{Brennecke}.

The electronic excitation is described now by the Hamiltonian $H_{ex} = \sum_nE_a\ B_n^{\dagger}B_n$, as before, for low excitations the operators $B_n$ are taken to be bosonic. The excitation-photon coupling Hamiltonian reads $H_I=\sum_{n}\left(f\ B_n^{\dagger}a+f^{\ast}\ a^{\dagger}B_n\right)$, where the coupling parameter is $f=-i\sqrt{\frac{h\nu_c\mu^2}{2\epsilon_0V}}$. The total Hamiltonian is given by
\begin{equation}
H = \sum_nE_a\ B_n^{\dagger}B_n+E_c\ a^{\dagger}a+\sum_{n}\left(f\ B_n^{\dagger}a+f^{\ast}\ a^{\dagger}B_n\right).
\end{equation}
We define the collective excitation operator $B_n=\frac{1}{\sqrt{N}}\ B\ ,\ B=\frac{1}{\sqrt{N}}\sum_nB_n$, to get
\begin{equation} \label{Hexcitation}
H = E_a\ B^{\dagger}B+E_c\ a^{\dagger}a+\bar{f}\ B^{\dagger}a+\bar{f}^{\ast}\ a^{\dagger}B,
\end{equation}
where $\bar{f}=-i\sqrt{\frac{h\nu_c\mu^2N}{2\epsilon_0V}}$. The Hamiltonian (\ref{Hexcitation}) is similar to the one for the coupled superradiant states and photons (\ref{Hexciton}), with the differences in the energies, $E_a$ and $E_{ex}$, and the coupling parameters. The energy shift is $E_a-E_{ex}=-2J_{\theta}\ \cos\left(\frac{\pi}{N+1}\right)$ and equal to about $1.35\times 10^{8}\ [Hz]$ for $N=10^{3}$ and $\theta=0$, which is a significant shift. The diagonalization of the Hamiltonian is exactly as before, and also the linear spectra calculations. For the case of resonance $E_c=E_a$, we have $|\bar{f}|/h=2.8\times 10^{7}\ [Hz]$. In figure (6) we compare between the vacuum Rabi splitting frequency, $\Omega_0=2|f|/h$, as a function of the atom number for the two cases. It is seen that for large atom number the vacuum Rabi splitting for independent atoms is larger than that of interacting atoms. The difference is about $5\times 10^{6}\ [Hz]$ for $N=10^{3}$ atoms. We conclude that the vacuum Rabi splitting is reduced by the dipole-dipole interactions.

\begin{figure}[h!]
\centerline{\epsfxsize=5.0cm \epsfbox{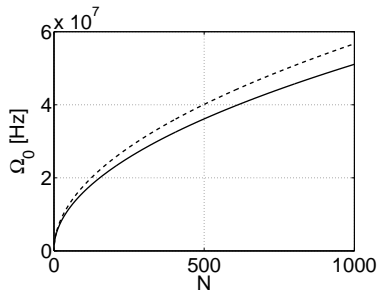}}
\caption{The vacuum Rabi splitting frequency $\Omega_0$ vs. the atom number $N$, for interacting (full line), and non-interacting (dashed line), optical lattice ultracold atoms. The difference increase with the atom number.}
\end{figure}

For more comparison we plot the generalized Rabi splitting for the two cases. We consider the case with $E_a=E_c$. Then for atoms without dipole-dipole interaction we have $\delta=0$, and then $\Omega=2|\bar{f}|/h$, which is a square root function of $N$, and $\theta$ independent. For the case with dipole-dipole interactions, we consider only the superradiant state, to get the generalized Rabi splitting, which is $\Omega =2\Delta/h$, where $\Delta$ defined previously, and which is $N$ and $\theta$ dependent. In figure (7a) we plot the generalized Rabi splitting as a function of $\theta$ for the two cases, and with $N=10^3$. It is clear that around $\theta=54.74^o$ the generalized Rabi splitting for interacting case is lower than the non-interacting case, where around $\theta=54.74^o$ the dipole-dipole interaction is zero, that is $J_{\theta}\approx 0$. In figure (7b) we plot the generalized Rabi splitting as a function of $N$ for the two cases, where for the interacting case we plot for the angles $\theta=0^o,\ 54.74^o,\ 90^o$.

\begin{figure}[h!]
\centerline{\epsfxsize=4.3cm \epsfbox{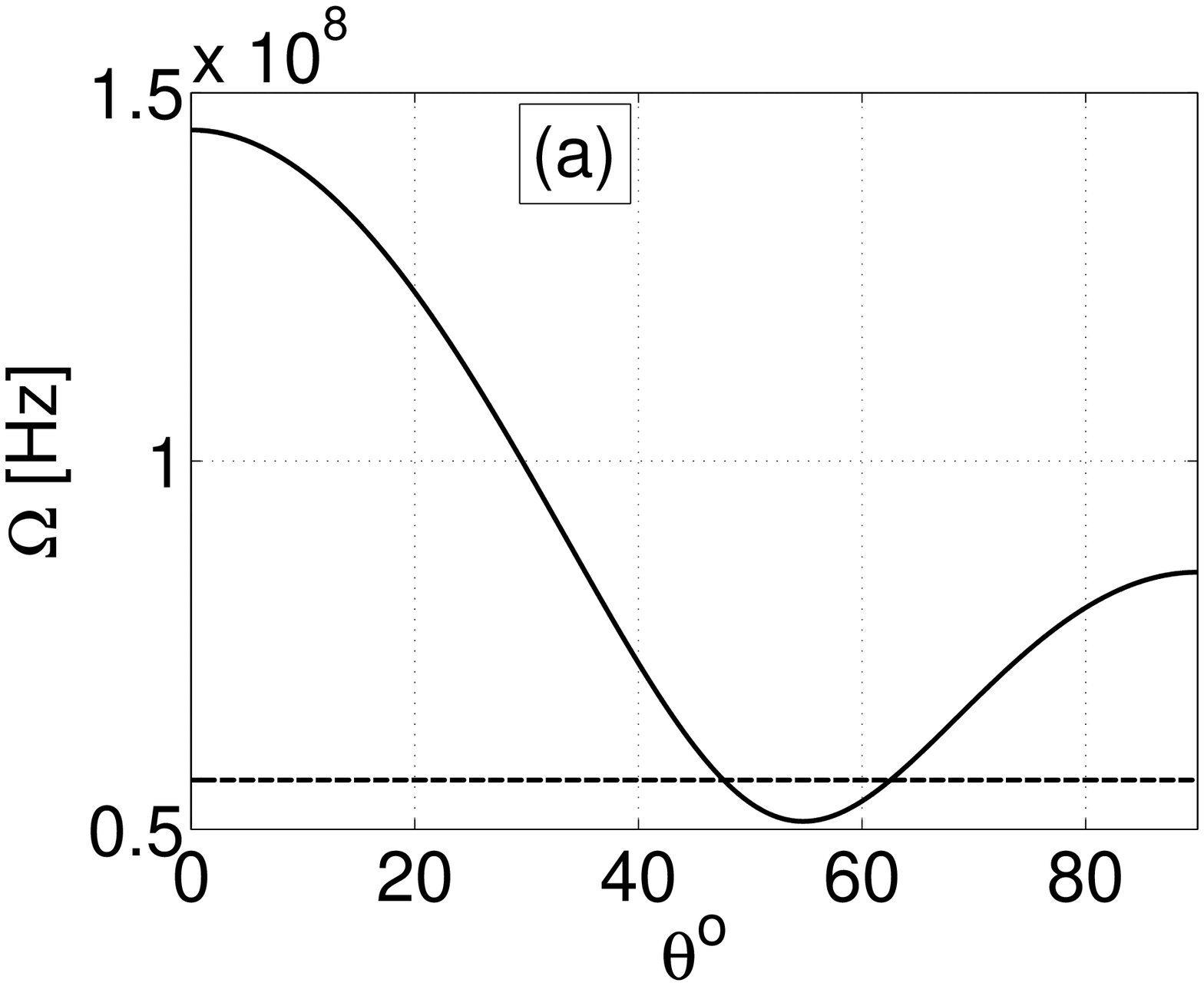}\epsfxsize=4.3cm \epsfbox{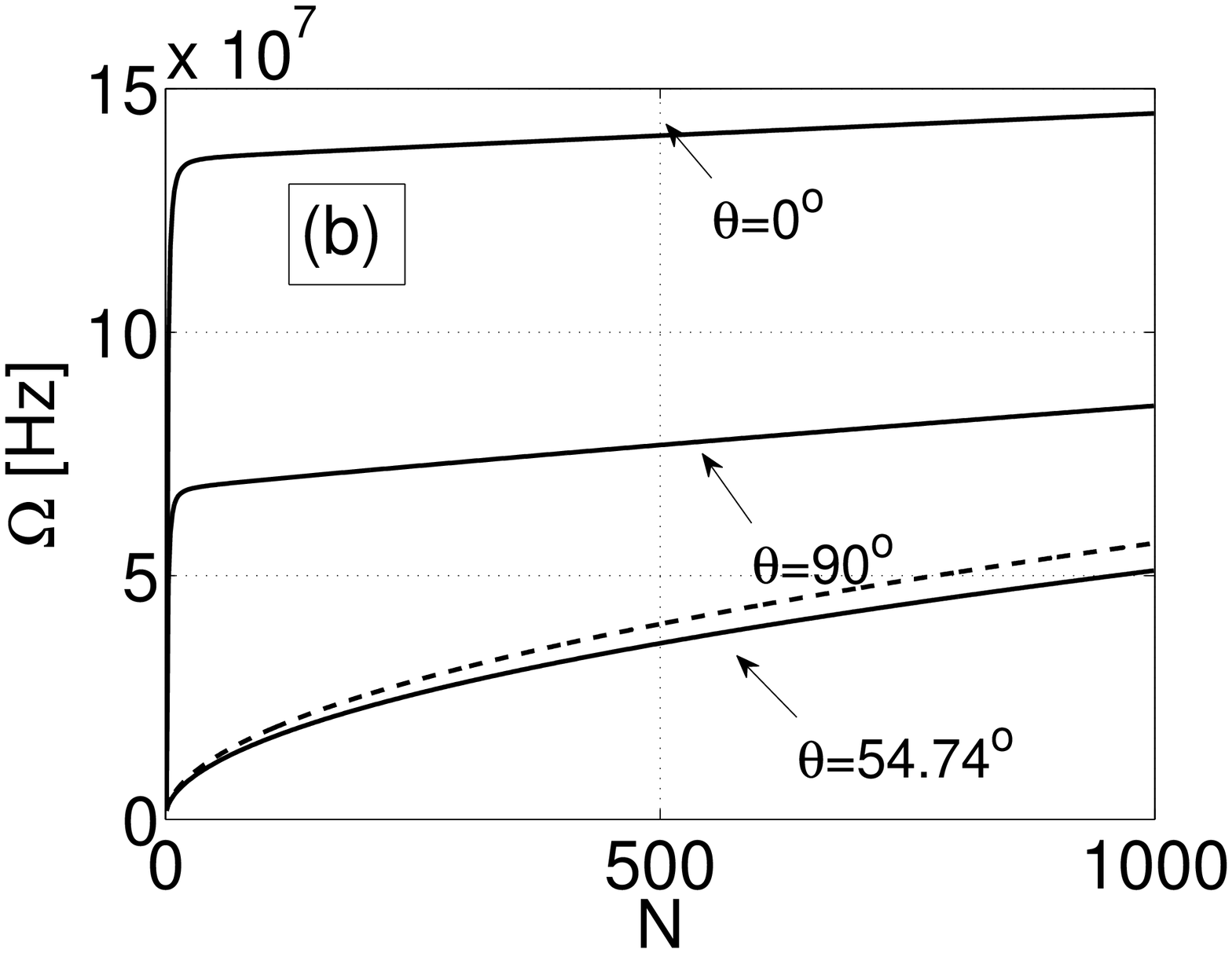}}
\caption{(a) The generalized Rabi splitting frequency $\Omega$ vs. $\theta$, for interacting case (full line), and the non-interacting case (dashed line), with $N=10^3$. The non-interacting case is $\theta$ independent. (b) The generalized Rabi splitting frequency $\Omega$ vs. $N$, for interacting case (full lines) at the angles $\theta=0^o,\ 54.74^o,\ 90^o$, and the non-interacting case (dashed line). Around $\theta=54.74^o$ the interacting case splitting is smaller than the non-interacting one.}
\end{figure}

In summary we explicitly calculated the coupled atom-field eigenmodes of a quantized cavity mode strongly coupled to collective excitations (excitons) of cold atoms in an optical lattice. Including resonance dipole-dipole interactions strongly influence the system eigenmodes and the electronic excitations form collective dark and bright states. Within a typical cavity only a single superradiant mode dominates the linear optical response forming a doublet of eigenstates. Note that analogous results also hold for any chain of electromagnetic active materials, e.g. a lattice of quantum dots within a cavity.

Acknowledgment: The work was supported by the Austrian Science Funds (FWF), via the project (P21101).

\end{document}